\documentclass[twoside]{article}
\usepackage{graphicx}  

\usepackage{amsmath}   
\begin{document}
\title{Lighthouses of Gravitational Wave Astronomy}
%
%
%
%
%
\author{Bernard~F.~Schutz\\ Max Planck Institute for Gravitational Physics\\ (Albert Einstein Institute)\\ 14476 Golm,  Germany\\ {\em and}\\ Department of Physics and Astronomy\\ University of Cardiff, Wales}
\date{}
%
%
%

\maketitle              

\begin{abstract}
Gravitational wave detectors capable of making astronomical 
observations could begin to operate within the next year, and 
over the next 10 years they will extend their reach out to cosmological
distances, culminating in the space mission
LISA. A prime target of these observatories will be binary systems, 
especially those whose orbits shrink measurably 
during an observation period. These systems are 
standard candles, and they offer independent ways of measuring cosmological 
parameters. LISA in particular 
could identify the epoch at which star formation began and, 
working with telescopes making electromagnetic observations, measure 
the Hubble flow at redshifts out to 4 or more with unprecedented accuracy.
\end{abstract}

\section{Introduction}
Gravitational wave interferometers now under construction at several 
locations around the world will soon begin making observations. Although 
their initial sensitivities will be marginal, a planned program of 
upgrades and technology development will make them powerful instruments 
of cosmology over the next decade. In 2011 the launch of the joint 
ESA-NASA gravitational wave observatory LISA will extend the reach 
of this form of astronomy to the entire observable universe. 

Astronomy has consistently proved itself to be full of surprises for 
observations in new wavebands, and this may well also be true for 
gravitational wave astronomy. Therefore it is a little dangerous to try to 
predic what these new detectors will observe, but 
it is useful to look ahead at this point. In particular it is not 
too early to consider and even to begin to plan the ways in which gravitational 
wave detectors and other astronomical telescopes could work together. 

For cosmology, one of the most interesting features of gravitational 
wave observations is that certain systems are standard candles: their 
distance can be inferred from their gravitational waveforms. These 
systems are chirping binaries, that is binary systems whose orbits 
shrink during the observations time because of the energy they lose
to gravitational waves. The change of the orbit raises the frequency
of the gravitational wave, producing a ``chirp'' waveform. 

In this review I will point out a number of ways in which gravitational wave 
observations of these chirping
waveforms, usually coupled with coordinated observations in electromagnetic 
wavebands,  can be used to provide cosmological information. For example, chirps from 
neutron-star binaries in the last few minutes before coalescence should be
observed frequently by advanced ground-based
detectors. These detectors can give astronomers advance notice of and rough 
positions for such inspiral events, and optical identifications of any afterglows produced by
the mergers of the neutron stars will sharpen the distance estimate 
made by the detectors. Redshifts to the afterglows can be used with these 
distance estimates to provide independent measurements of the Hubble constant 
to accuracies of a few percent, and of the acceleration of the universe out to 
redshifts of order 0.2. Chirps from mergers of stellar-mass black holes could, even
in the absence of electromagnetic counterparts, provide estimates of the cosmological
acceleration out to redshifts of order 1.

Chirps from very massive ($1000\,M_\odot$)
black hole binaries that might have formed in the first epoch of star formation, observed 
by LISA, could be used to determine when star formation began. Chirps from 
the coalescences of massive black holes in galactic centres, again observed by LISA, 
could measure the cosmological deceleration out to redshifts of 4 or more, 
provided that electromagnetic observations can pin down the cluster of
galaxies in which a coalescence occurred. This will be a real challenge
to astronomy but it could have an immense payoff. 

Before discussing these possibilities, I begin this article with two
background sections. The first reviews  
gravitational wave astronomy, particularly
emphasising the ways in which gravitational wave observations differ
in concept and information content from electromagnetic observations, and 
outlining the development 
of detectors and the timetable on which sensitivity improvements can 
be expected. The second iss an introduction to chirp waveforms and the kind
of information they carry. These two sections prepare for the subsequent discussions
of how cosmological information can be extracted from gravitational wave 
observations.

\section{Gravitational wave observing}
The principles of interferometric detectors and their current 
development are reviewed in a number of places in the 
literature~\cite{saulson,hough}. In particular, \cite{hough} contains 
references to the recent literature. Several accessible textbooks 
\cite{MTW,firstcourse} review the principles of gravitational radiation, and two recent 
encyclopedia articles also address these issues~\cite{schutzwill,schutzappphys}.
What follows here is a brief introduction to these subjects.
\subsection{Action of waves on a detector}
\begin{figure}[!b]
\begin{center}
\includegraphics[width=\textwidth]{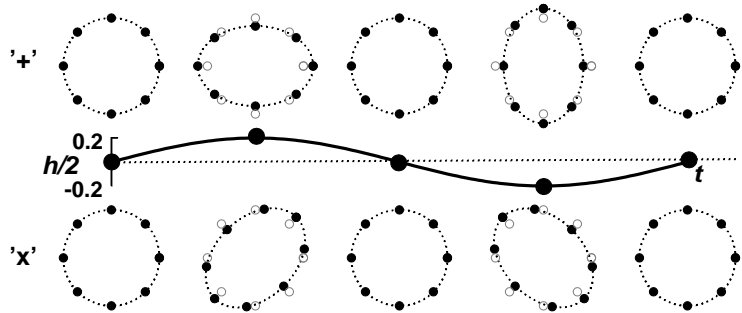}
\end{center}
\caption[]{Two independent polarisations of a plane gravitational wave
are illustrated by their actions on a ring of free particles in empty 
space. The waves act transversely, so in this figure the waves approach
perpendicular to the paper. The waves distort the ring 
into ellipses with alternating major 
and minor axes. The two polarizations are orthogonal because the ellipses
are rotated by $45^\circ$ with respect to one another. The action of 
any plane wave is a superposition of these two polarizations. The waves act
through tidal forces, so the stretching is 
proportional to the size of the ring: a given gravitational wave will
produce twice the relative displacement in a ring twice the size. The 
waveform in the centre shows the size of the strain illustrated here, 
which is defined as half of the amplitude $h$ of the wave. (This is 
much larger than we expect, of course.) The 
distortions produced by a given wave mimic the motions in the source 
of the wave as projected onto the plane of the sky as seen from the 
ring of particles. For example, if the source contains stars moving back and forth 
along the $x$-axis, then the wave will produce a similar motion in the 
ring (top line).}
\label{schutz_fig:polarization}
\end{figure}

Figure~\ref{schutz_fig:polarization} shows how gravitational waves act on 
a ring of free particles. The action is by tidal forces carried by 
the waves, which distort the ring in directions transverse to the 
direction of propagation of the wave. Because of the equivalence 
principle, the overall acceleration of the ring produces no local 
effects; the only measurable effects are in the relative distortions.
Therefore the linear displacements of these 
distortions are proportional to the size of the ring: the larger
the ring, the larger the displacement. 

The figure can be viewed as an elementary detector. By sensing the 
relative displacements of particles on the ring, one can measure
the wave. This is exactly how interferometric detectors work. They 
use laser interferometry to measure changes in the relative distances
between the central particle of a ring and two particles in orthogonal 
directions along the circumference of the ring. The particles are 
mirrors in the interferometer that are free to move along the 
direction of the displacement. 

Older, solid-mass detectors, called bar detectors, use the 
stretching along one diameter of the ring. The restoring forces of the 
solid material means that the response to the wave is more complicated
than in an interferometer, but the principle is the same.

All proposed gravitational wave detectors are linearly polarized. 
To measure the polarization of a wave requires either that several
detectors make measurements or that the wave lasts long enough so
that the motion of a detector carried by the Earth or in a space orbit
changes the projected polarization of the detector, allowing it to 
measure two independent polarizations.

The waveforms in Fig.~\ref{schutz_fig:polarization} 
have a simple relationship to the mass motions 
in the source of the gravitational wave~\cite{como}. 
They mimic any oscillating 
motions in the source, as projected on the plane of the sky as seen
by the ring. If the motions are all along one line, then the 
polarization ellipse will have its alternating major/minor axes
along that line. If the motions are circular, then the wave will
have circular polarization, which is a linear combination of the two
polarizations with a phase shift of 90 degrees. Thus, measuring the 
polarization of a gravitational wave allows one to make direct 
inferences about the source, such as measuring the angle of inclination
of the orbital plane of a binary system.

\subsection{Planned ground-based detectors}
I will focus here on the planned interferometers, which have the 
biggest potential for astronomical and particularly for cosmological
observing. There are four instruments now being built~\cite{hough} that should 
reach the target sensitivity of {\em first-generation} detectors:
to measure $h\sim 10^{-21}$. This is a threshold that has been the goal 
of detector development for decades: it is the largest amplitude
that could reasonably be expected from sources that might be 
observable in one year. Observing at this level gives no guarantees of
detections: Nature has to cooperate by providing strong sources. These
first-generation detectors are the first step along a planned sequence
of sensitivity improvements that will produced essentially guaranteed 
detections by the end of this decade.

Interferometers use light to compare the lengths of the two arms. The 
fundamental limit on sensitivity is the amount of light, since quantum 
uncertainties in the arrival times of photons produce a stochastic
noise called shot noise, which is less important when there are more
photons. However, the main technical challenge to these detectors is
to eliminate low-frequency noise from external vibrations and from 
internal thermal vibrations of the components~\cite{hough}. These noise sources 
will set a lower-frequency limit of about 40~Hz on LIGO and GEO. The 
VIRGO detector is investing more effort in controlling vibration noise, 
and will have some sensitivity even at 20~Hz. All detectors go up to
a few kHz before shot noise limits their sensitivity.

The largest and most ambitious project is LIGO, an American project 
building two 4-km detectors, one at Hanford (WA) and the other at
Livingston (LA). The Hanford detector also contains a 2-km instrument
for local coincidence and anti-coincidence observing. LIGO is now 
successfully doing interferometry, and is improving its sensitivity and 
reliability. LIGO may begin taking data at its planned sensitivity within
the next year. 

The next-largest instrument under construction is VIRGO, near Pisa. It is 
a cooperation between France and Italy. The timescale for operation of 
this 3-km instrument is about a year behind LIGO.

In Germany the GEO project is building a smaller detector, GEO600, with
 600-m arms, that will nevertheless have a similar sensitivity to the 
LIGO instruments, and which is on the same timescale as LIGO. It 
achieves this sensitivity by using more advanced optical and mechanical 
technology. This technology will be transferred to LIGO and VIRGO when 
they are ready for upgrades to higher sensitivity.

GEO600 and LIGO are planning a joint test data run in December 2001, and the 
two projects have in fact signed a strong data-sharing and data-analysis
MOU, providing for joint publication of all results. Other projects
have been invited to join in this agreement. 

The second-generation instruments will be upgrades of LIGO and VIRGO, 
which could be in place by 2007, plus a proposal in Japan that is not
yet funded. These will improve the first-generation sensitivity by 
a factor of 10 in amplitude, and they will push the observing frequency
limit down to perhaps 10~Hz. Scientists are beginning to design radical new 
technology for the third generation, envisioning yet a further step 
by a factor of 10, and a further broadening of the observing 
frequency window. It is possible that VIRGO and GEO will cooperate on 
a joint proposal for a new third-generation detector in Europe.

\subsection{LISA, the first space-based detector}
Ground-based detectors will never have sufficient sensitivity to do useful
work below about 1~Hz, because gravity noise generated by moving masses on 
the Earth will be larger in amplitude than expected gravitational waves.
Since gravity cannot be screened, the only solution is to put the detector
into space. This is the justification for LISA, which is planned for launch
in 2011. Unlike the ground-based detectors, LISA will observe many of 
its sources with extremely high signal-to-noise ratio. LISA is likely 
to be the first of a sequence of space-based detectors over the next 
few decades. LISA could have a mission lifetime of up to 10 years. 
The state of development was reviewed recently in the proceedings of
the Third International LISA Symposium~\cite{lisa3}.

LISA began in 1993, when an American group led by P Bender of JILA, which 
had been studying space-based detectors for some time, encouraged  
a group of Europeans, largely in the GEO and VIRGO projects, to 
propose a detector for the ESA M3 mission opportunity. The mission was
not selected because it was too expensive for this medium-mission limit, 
but the scientific potential was regarded so highly that the group was
encouraged to propose for the Horizon 2000+ Cornerstone selection in 
1995. The present design, based on a triangular three-armed 
interferometer, with a detailed plan for the optics and sensing 
needed, matured for that proposal. 

LISA was indeed selected as a Cornerstone, but still the costs were 
troubling. A redesign by the European LISA team, cooperating with
JPL and Bender's group at JILA, produced the current baseline 
design using three spacecraft, and seemed to be affordable. The 
project meanwhile gained considerable interest among astronomers, 
who were coming to the conclusion that the giant black holes that
LISA could observe were ubiquitous in the centres of galaxies. 
This led to efforts to bring NASA into the project to share costs.

Earlier this year (2001), ESA and NASA exchanged letters of agreement
to share the project equally, and ESA invited NASA to contribute to a technology
demonstration mission called SMART~2, due for launch in 2006. The 
technology of LISA is a fascinating subject in itself, which there is 
no room for here. The two agencies have formed a joint LISA International
Science Team (LIST), that will organise the community. It has two chairs, 
T Prince (Caltech and JPL) and K Danzmann (of the new branch of 
the Albert Einstein Institute in  Hannover). Theorists and astronomers 
who want to contribute to the science required before LISA's launch 
are welcomed to join in the projects being encouraged by the LIST's
Sources and Sensitivities Working Group, jointly chaired by S Phinney
(Caltech) and the present author. 

\begin{figure}[!b]
\begin{center}
\includegraphics[width=.6\textwidth,angle=90]{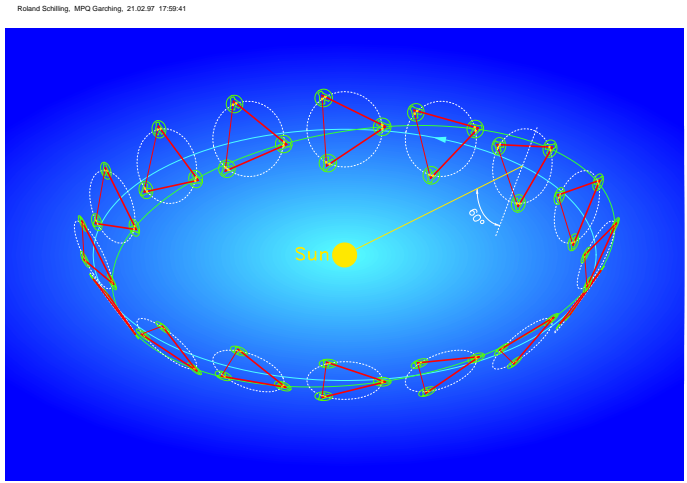}
\end{center}
\caption[]{LISA's configuration remains roughly equilateral and 
turns about an axis perpendicular to the triangle's plane, 
while the triangle remains in a plane
tilted $60^\circ$ to the ecliptic as the spacecraft orbit the Sun 
in one year.}
\label{schutz_fig:lisa}
\end{figure}

As mentioned above, LISA is a three-armed detector. The roughly 
equilateral triangle maintains its shape as the three spacecraft 
follow their independent orbits around the Sun. The triangle 
lies in a plane tilted $60^\circ$ to the ecliptic, and is situated
about $20^\circ$ behind the Earth in its orbit.  The arm-length of 
5 million kilometres permits good sensitivity below 0.1\,Hz. The
lower limit on LISA's frequency window is about 0.1\,mHz, where 
perturbations due to fluctuations in the solar radiation pressure
dominate.  Figure~\ref{schutz_fig:lisa} shows the way the 
location and orientation of LISA change during a year.

\subsection{Principles of observation with gravitational waves}
Gravitational wave observing is rather different from observing in
the electromagnetic spectrum. This is partly because detectors cannot 
be pointed: they are simple quadrupoles with broad response patterns 
on the sky. And it is partly because detectors register
the waves coherently, following the oscillations of phase. By contrast,
most electromagnetic detection is bolometric, registering the energy
and not the phase. Even radio interferometry, which uses phase at an 
early stage, eventually rectifies the signal and records the energy in 
the fringes. Detectable electromagnetic waves simply oscillate too fast
to be recorded, and their phase in any case does not necessarily contain
important information.

The gravitational wave phase oscillated at kHz frequencies or lower,
and directly reflects the mass motions in the source. Almost all the 
useful information in a signal is in the phase $\phi(t)$. This has 
several implications that are not immediately obvious to astronomers
used to electromagnetic observing.

First, spectroscopy and polarimetry are automatic. The detectors are
linearly polarised, and spectroscopy is nothing more than taking the 
Fourier transform of the detected signal.

Second, detecting a gravitational wave usually means being able to 
measure several parameters, such as the masses of the systems, that 
are encoded in the phase and frequency information. Moreover, as 
explained earlier, the polarisation information can be used to infer
source orientations.

Third, observing by multiple detectors brings great benefits, particularly 
in angular resolution, as well as in confidence when signals are weak.
The angular resolution improvements are analogous to what happens in 
radio interferometry. Even single detectors can achieve this if they
observe a continuous source long enough to take advantage of the changing
detector position; in this case such a detector effectively does 
aperture synthesis by itself.

Fourth, data analysis on computers plays a crucial role in detection. 
 The optimum detection strategy in gravitational wave 
observing is to employ matched filtering~\cite{kip300}. This means comparing the 
observed phase $\phi_{\text{true}}(t)$ with the expected phase of a 
template signal $\phi_{\text{template}}(t)$. If the two match well enough, 
which usually means that $|\phi_{\text{true}}(t)-\phi_{\text{true}}(t)|< 1\,\text{radian}$ over the signal duration, then one is close to optimal. Since
the templates depend on parameters, which describe physical properties like 
sky location, source masses, orbit spindown, and other effects, 
a search for signals typically involves many repeated comparisons with 
slightly different templates. Then the availability of computer power 
(or the lack of it)  
can limit the sensitivity of an observation. This sometimes happens 
for other kinds of observing, for example the search for binary radio
pulsars, where the parameter space that must be searched for signals is
non-trivial in size.

Fifth, gravitational wave astronomers always speak about detecting 
{\em amplitudes}, not energy. This means their signal-to-noise ratios are
 the square-roots of energy or flux-based signal-to-noise measures.
So if a gravitational wave observation with LISA can reach a signal-to-noise
ratio of $10^4$, then this should be compared with an optical 
observation with a ratio of $10^8$: one photon of background for each
$10^8$ photons from the source! This is suggestive of how much detailed 
information is potentially extractable from LISA observations.

There are many analogies between long-duration gravitational wave observations
and radio observations of pulsars, in that radio observations are coherent
as regards the pulse period itself. This is similar to the gravitational 
wave period of waves from the same pulsar, so many issues are the 
same. For example, gravitational wave positions will be at the same
accuracy level as radio positions, around the arcsecond mark.

\subsection{Angular positions}
Once a source of gravitational waves has been detected, the most important 
information that the observation can
produce is, of course, the location of the source on the sky. The 
accuracy of angular positions will be the crucial step in identifying 
sources and opening them for study by electromagnetic observation. 

Since, as 
we remarked above, the pointing accuracy of an individual detector is 
poor over a short observation time, the position of the source must use
more information than the instantaneous response of a single detector. The 
accuracy of locating a short burst comes entirely from the simultaneous 
observation of the event  by several detectors. The accuracy of a long-duration
observation is achieved, as mentioned above, by aperture synthesis.

A short burst may be defined as one in which the acceleration of the detector 
during the observation does not produce an overall phase-shift of the 
wave-form by more than one radian. The overall motion of the detector
produces a constant (and usually unobservable) Doppler effect, but the 
acceleration of the detector distorts the waveform, and this can tell us
where the wave came from. If the detector acceleration is $\vec{a}$ and 
the wave-vector of the radiation from the source is $\vec{k}$, then during
an observation lasting a time $T$, short enough to regard the acceleration 
as constant, the phase-shift induced by the acceleration is
\[\Delta\phi_{\text{accel}}=\frac{1}{2}(\vec{a}\cdot\vec{k})T^2.\]
The condition that this should be less than 1 amounts, for a typical 
value of the gravitational wave frequency $f_{\text{gw}}$ and for the 
acceleration produced by the rotation of the Earth, this sets a limit 
on the time of observation of
\begin{equation}\label{schutz_eqn:burst_criterion}
T_{\text{burst}} <  56\left(\frac{f_{\text{gw}}}{1\,\text{kHz}}\right)^{-1/2}\,\text{min}.
\end{equation}

For such bursts, the position must be triangulated by using the arrival 
times of the waves at several detectors. This uses the detectors as an 
interferometer array, and the pointing accuracy is the diffraction limit, 
roughly the wavelength of the waves divided by the detector spacing. Within 
this, a source with strong signal-to-noise ratio can be located more 
accurately. If SNR is the amplitude signal-to-noise ratio for a particular
observation, then for detectors with a baseline between Europe and the USA, 
say $10^4$~km, the accuracy is~\cite{cutlerflanagan}
\begin{equation}\label{schutz_eqn:burst_pointing}
\Delta\theta_{\text{burst}}\sim\frac{2^\circ}{\text{SNR}} \left(\frac{f_{\text{gw}}}{1\,\text{kHz}}\right)^{-1}.
\end{equation}
A confidence limit for detection will be something like $\text{SNR}>5$, so 
that any detected source might be triangulated to better than half a degree. 
A strong source with $\text{SNR}=20$ could be located to within 5 to 10 arcminutes. 

These are overly optimistic numbers, however, because there is covariance
with other observational errors. If there is an error in determining the 
polarisation, then this could masquerade as a delay or advance in the signal 
by up to half a cycle. What is more, the diffraction limit applies only 
if there are enough detectors to determine the polarisation. With three 
detectors there are two possible solutions to the location on the sky. The 
ambiguity is resolved only with four or more detectors. Moreover, if 
the detectors are unusually well aligned, then they do not determine 
polarisation as well. Unfortunately this is the case: the LIGO detectors, 
in the interests of ensuring that they should see nearly identical 
responses, are very well aligned, so they do not contribute much to 
a position determination. The result is that real position determinations 
for bursts might be five to ten times worse than the numbers quoted above. 

This situation could be significantly improved if a detector is built in 
Japan, with its long baseline to the others. Such a detector would 
improve both the detection rate and the position determinations by  
a factor between two and four.

A long-duration source permits a single detector to determine the position
by using the phase-modulation and time-dependent polarisation projection
to measure both the polarisation and position jointly. The best case 
is when the source lasts for a year, so that the detector synthesises 
a telescope with an aperture of 2\,AU. The diffraction-limited position 
accuracy improves on the above to
\begin{equation}\label{schutz_eqn:cw_pointing}
\Delta\theta_{\text{continuous}}\sim\frac{0.5\,\text{arcsec}}{\text{SNR}} \left(\frac{f_{\text{gw}}}{1\,\text{kHz}}\right)^{-1}.
\end{equation}
Again, this is a little optimistic because polarisation errors can add up to 
a cycle to the waveform. But the pointing accuracy for ground-based detectors
observing pulsars, for example, is very good. However, LISA will observe
in the mHz region, which degrades its position accuracy. This is compensated
somewhat by the large SNR, so that the result is a resolution accuracy between 
10 arcminutes and 10 degrees, depending on how strong the source is. We will 
come back to the importance of these errors in the next section.

\subsection{Amplitude estimates}
The use of chirping binaries as standard candles depends on being able 
to measure the amplitude of their radiation accurately. In principle, 
this is just what the SNR measures, so the amplitude error would be of
order $1/\text{SNR}$. But there is a strong covariance with the position 
error, since the antenna pattern of the detectors is broad. Roughly speaking, 
a position error of $\Delta\theta$ measured in radians produces a relative
change in the sensitivity of the 
detector with respect to the source by a comparable amount. This will 
result in a wrong determination of the amplitude. So a good rule of thumb for 
amplitude errors is:
\begin{equation}\label{schutz_eqn:amplitude}
\Delta h = \max(1/\text{SNR},\Delta\theta).
\end{equation}
If the only observations of the event are from gravitational wave detectors, 
then $\Delta\theta$ must be inferred from the equations above. But if 
the event can be identified by electromagnetic observations, then the position 
accuracy can be much improved, and with it the amplitude accuracy. This 
is particularly the case for LISA, where the SNR could be as high as 
$10^4$, but the intrinsic position accuracy could be as bad as  0.2 radians~\cite{cutler}. 
{\em The 
astronomical return from LISA observations can be greatly improved by 
coordinated electromagnetic observations.}

\section{Chirping Binaries}
\subsection{Distance determination: the standard candle}
When LISA or a network of ground-based detectors observes a binary system, 
then they can determine the angular position, amplitude, and angle of 
inclination of the binary orbit, as described above. For simplicity 
let us now assume that the orbit is circular, although what we describe
can be extended to elliptical orbits.

There is a remarkable coincidence in the radiation from binary systems, 
in that both the amplitude $h$ of the radiation and the rate of change
of the frequency of the radiation $df_{\text{gw}}/dt$ depend on the 
masses of the two stars only through exactly the same combination, 
which is called the {\em chirp mass} {\cal M} of the system. If 
the two stars have masses $m_1$ and $m_2$, with associated reduced 
mass $\mu$ and total mass $M_T$, then the chirp mass is defined by
\begin{equation}\label{schutz_eqn:chirpmass}
{\mathcal M} := \mu^{3/5}M_T^{2/5} = (m_1m_2)^{3/5}(m_1+m_2)^{-1/5}.
\end{equation}
This was first pointed out by the present author~\cite{schutzhub}, who suggested how 
this could be used to measure the luminosity distance $d_L$ to any binary 
system that chirped, that is whose $df_{\text{gw}}/dt$ could be 
measured. 

One way to see how this can be done is to consider the formula for
the SNR of an observation using a filter that has been perfectly matched
to the incoming signal, in polarisation and chirp mass. 
We consider only the radiation from the orbit, not from the later
coalescence event. This underestimates the SNR, but it has the 
advantage that the orbit is fully understood and its SNR can 
be characterised, while the radiation to be expected from coalescence 
is not yet known.  Then the 
SNR can be written in the following way for a burst chirp ~\cite{chernofffinn}, that is 
a chirp that lasts less than the time given in 
(\ref{schutz_eqn:burst_criterion}):
\begin{equation}\label{schutz_eqn:chirpSNR}
\text{SNR}= 8\Theta\frac{r_0}{d_L}\left(\frac{{\mathcal M}}{1.2 M_{\odot}}\right)^{5/6}\zeta(f_{\text{max}}).
\end{equation}
The following terms enter this equation:
\begin{itemize}
\item $\Theta$ is a factor that depends on the projection of 
the polarisation of the wave on the antenna pattern, so it is a function of 
the orientation of the binary relative to the detector. This is 
measurable from the polarization and direction information.
\item $r_0$ is the {\em range} of the detector for this kind of 
observation, that is a distance that depends on the sensitivity of the 
detector.  It is a function only of the detector.
\item $d_L$ is the luminosity distance to the source, and is what we 
want to determine from the observation.
\item ${\mathcal M}$ is the chirp mass, and is determined from the 
observed rate of change of the frequency of the chirp.
\item $\zeta$ is a number that depends weakly on $\mathcal M$, taking 
account of the fact that massive chirping systems reach coalescence and 
hence their maximum frequency at a lower frequency than less massive ones do, 
so the response of the detector to them is a little different. This 
is clearly also a function of the detector, but it is known once $\mathcal M$ 
has been determined. 
\end{itemize}
From this list it is clear that all the numbers in this equation, including 
the value of SNR, are determined either by the detector or by the 
observation of the signal, except for $d_L$. This is therefore the unknown 
that can be solved for. The result is that {\em observations of the 
radiation from the orbit of a chirping binary determine its luminosity 
distance.}

\subsection{Which binaries chirp?}
\begin{figure}[!b]
\begin{center}
\includegraphics[width=\textwidth]{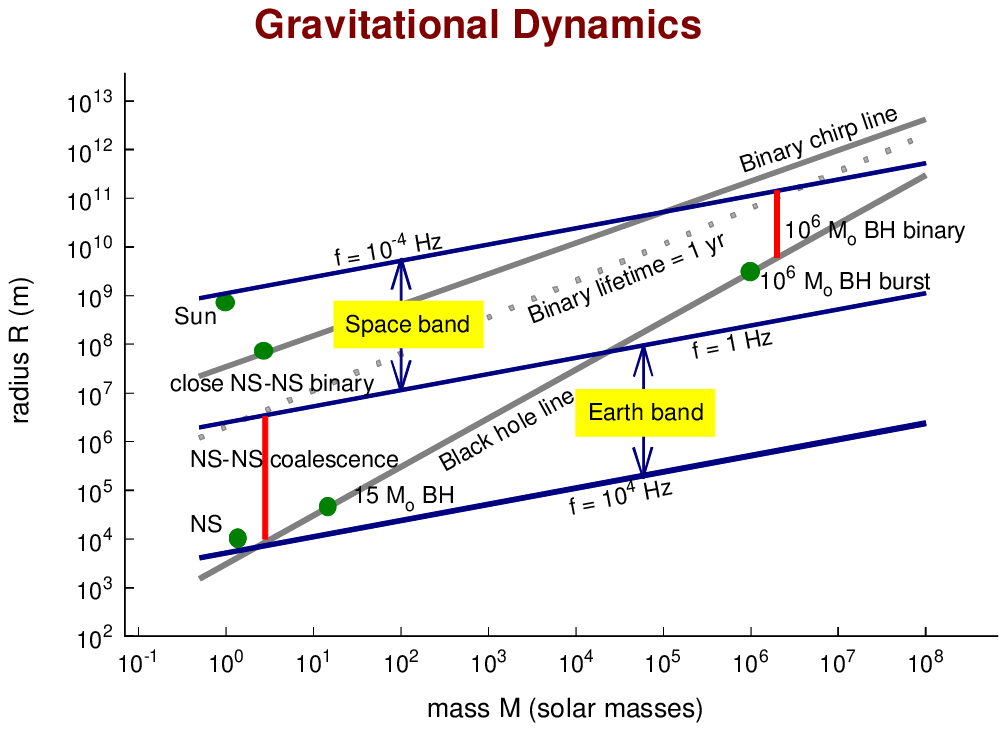}
\end{center}
\caption[]{A rough classification of binary systems according to 
their total mass (horizontal axis) and size (vertical axis). The three solid
lines with slope 1/3 in this log-log chart show systems whose natural
frequencies $(\pi G\varrho)^{1/2}$  are $10^{-4}$\,Hz, 1\,Hz, and $10^4$\,Hz, 
respectively. These divide the chart into systems in the frequency region
that one can observe from space and the ground. The line with slope 1 is
the black-hole line: systems below this line do not exist. The dotted line 
is the line below which systems will coalesce within one year. The solid 
chirp line, on the other hand, delimits systems that can be observed to 
change their frequency in one year. Notice that all binaries observable 
from the ground will coalesce within one year, while most binaries 
observable by LISA with masses above about 1000 solar masses will chirp.}
\label{schutz_fig:chirps}
\end{figure}

The expression for the rate of change of the frequency of radiation from 
a binary alluded to above can be formulated to give a characteristic 
time called the {\em chirp time} $\tau_{\text{gw}} = (df_{\text{gw}}/dt)/f_{\text{gw}}$. Here are some useful ways to calculate this chirp time for various 
interesting systems:
\begin{eqnarray}
\tau_{\text{gw}} &=& 200 \left(\frac{f}{20\,\text{Hz}}\right)^{-8/3}
\left(\frac{{\mathcal M}}{1.2M_\odot}\right)^{-5/3}\,\text{s},\label{schutz_eqn:chirptimeone}\\
&=& 44 \left(\frac{f}{20\,\text{Hz}}\right)^{-8/3}
\left(\frac{{\mathcal M}}{3.0M_\odot}\right)^{-5/3}\,\text{s},\label{schutz_eqn:chirptimetwo}\\
&=& 44 \left(\frac{f}{10^{-4}\,\text{Hz}}\right)^{-8/3}
\left(\frac{{\mathcal M}}{10^6 M_\odot}\right)^{-5/3}\,\text{s},\label{schutz_eqn:chirptimethree}\\
&=& 4\times 10^5 \left(\frac{f}{3\,\text{mHz}}\right)^{-8/3}
\left(\frac{{\mathcal M}}{0.5M_\odot}\right)^{-5/3}\,\text{s}.\label{schutz_eqn:chirptimefour}
\end{eqnarray}
This list shows how long one must wait for a system to change its frequency 
substantially, say by a factor of about two. However, one does not have  
to wait that long to see a system chirp. All that one requires is that the
system change its frequency by the frequency resolution of the 
observation, which is $1/T_{\text{obs}}$ for an observation of duration
$T_{\text{obs}}$. Figure~\ref{schutz_fig:chirps} shows the systems that 
chirp and those that coalesce within a one-year observing time.

From these times, it is possible also to calculate the population 
statistics of chirping systems if they are created at a fixed steady
rate $R$ and die away through coalescence. Then the steady-state 
population of systems radiating gravitational waves with a frequency
larger than any given $f_{\text{gw}}$ is 
\begin{equation}\label{schutz_eqn:population}
N(>f) = 7000\left(\frac{f_{\text{gw}}}{0.1\text{mHz}}\right)^{-8/3}
\left(\frac{{\mathcal M}}{1.2M_\odot}\right)^{-5/3}
\left(\frac{R}{10^{-5}\text{yr}^{-1}}\right).
\end{equation}
For example, in the Galaxy, systems like the Hulse-Taylor pulsar are expected to form 
once every $40^5$\,yr or so, so this means that there should be thousands
of such systems within the LISA waveband. LISA should 
certainly see many if not most of them, depending on whether they are 
obscured by radiation from the far more common white-dwarf binary systems.

Another interesting number is to try to estimate the population of 
binaries of $1000 M_\odot$ that might be observable by LISA. LISA should 
see any chirping system in its waveband anywhere in the Universe. In 
the first generation of stars, suppose that 10\% of the present 
stellar mass of the universe went into 
such stars during a time lasting about $10^9$ years. Then there would have 
been some $10^{18}$ such stars that formed. Most of them may well have evolved
into black holes~\cite{heger}, and the gravitational collapse event might not have 
disrupted the binary system. Any such binaries with a local gravitational 
wave frequency above 6\,mHz would, by the above formulas, coalesce within 
about 1 year of our time, so LISA could follow the event most of the 
way to coalescence. 

Suppose a fraction $\eta$ of such systems
formed binaries that could coalesce in the first $10^9$ years. Then 
the rate of formation of such binaries was $10^9\eta$ per year. By 
(\ref{schutz_eqn:population}), the number of such systems that LISA could 
follow to coalescence in one year of observing is about $1700\eta$. So 
if the efficiency of formation of these binaries is better than one tenth
of one percent, then LISA would be able to detect a few. If the efficiency
is better than 1\%, then LISA would have of order 200 events during 
a ten-year mission lifetime, and the upper limit on the luminosity 
distances to these events would signal the onset of this first generation 
of star formation.

\section{Cosmology with Ground-Based Detectors}
The use of chirping binaries as standard candles to discover cosmological 
information has been studied by a number of 
authors~\cite{schutzhub,markovic,chfinn,finn,turner}.
They have pointed out that there is a variety of methods to  avoid the problem 
of identifying the galaxy in which the chirp occurred, and still extract 
cosmological information. I begin here, however, with the expectation that 
chirps may produce gamma-ray bursts, and in any case should certainly produce
optical/radio/X-ray displays of some kind, which I will call ``afterglows''.  These will help
the identification of the event. After this discussion, I return to the subject
of identifying black hole coalescences, which should have no optical counterpart.

\subsection{Afterglow cosmology}
As we have noted above, the second-generation ground-based detectors
should have ranges so that confident detections of coalescences of 
binary neutron stars can be made out to 400\,Mpc or so. Tens of events 
per year are to be expected~\cite{lorimer}. However, the 
errors in position determination are rather large, so that if no 
other information is available then it will be difficult to determine 
the galaxies in which the event occurred. 

The situation will be dramatically different if such events lead to 
gamma-ray bursts, or indeed if they lead to any other kind of 
transient event that leaves behind an afterglow. This seems very likely. 
From the identification of the event by electromagnetic detection of 
the afterglow, the position can be determined with very small errors
and the redshift of the galaxy can be measured. Then the 
luminosity distance can be determined within errors given just by 
the SNR, which could be of order 10. Thus, each event leads to 
a value for the Hubble constant accurate to 10\%. With 50 events 
over a few years of observing, the statistical errors could go down 
to a few percent. Although the Hubble constant should be known to 
this accuracy by other means by the time second-generation detectors
operate, this method will be an important check on the systematics of 
other determinations. 

Being able to give advance notice of a burst event by gravitational 
waves will also be a valuable contribution of these detectors. The 
gravitational wave signal will precede even the gamma-ray burst, 
and the detector scientists plan to build early-warning alert systems
so that notice of potential events could be available to cooperating
astronomers within seconds of the gravitational wave observation. In 
advanced detectors, the inspiral signal for two neutron stars could  
last several minutes, and there could be enough signal after the first
minute to predict the coalescence event. In this case, astronomers 
would have a minute or so notice to begin observing before the 
actual coalescence event even occurred.

The association of gravitational wave events with afterglows will 
of course immensely help modelling of gamma-ray bursts, and it will 
also allow estimates of the beaming fraction. Gravitational wave 
emission by these systems is much more isotropic than the gamma 
radiation, and therefore detectors should give a fair sample of 
all coalescing systems within their range.

It is possible that gamma bursts are associated more strongly with 
coalescences between neutron stars and black holes than between double 
neutron stars. If this is the case, the second-generation 
detectors will have a longer range, out to redshifts of order 0.3. 
This will make their ability to do cosmology much more interesting, 
and measurements of the local acceleration of the universe to $\pm10$\% would 
be possible over 5 years. 

All of these numbers improve by factors of 2 or more 
if a further detector is added to the network, say in Japan.

\subsection{Cosmology with binary black hole observations from the ground}
It is possible that the event rate for coalescence of binary black holes
of stellar mass will be comparable to that for binary neutron stars. 
Binaries are less likely to be disrupted by black-hole formation than 
by neutron-star formation because less mass is lost. And globular 
clusters seem to be efficient factories for black-hole binaries~\cite{port}. 
It may happen, then, that the first events detected by ground-based
instruments will be black-hole coalescences. And if that is the case, 
then second-generation detectors may see many tens of such events out
to redshifts of order 1. Over this distance, it is no longer appropriate
to speak of the Hubble constant or the deceleration parameter, since
these are just terms in the Taylor expansion of the recession velocity.
The observed acceleration of the universe makes such a local approximation
inadequate. The goal over cosmological distances is to sample the function 
$z(d_L)$, the Hubble flow, over as large a range of values of $d_L$ as possible.

Unfortunately, distant as these black-hole events are, they 
do not produce afterglows or other electromagnetically 
detectable counterparts from which a redshift can be measured. 
To circumvent this, I have proposed 
a statistical method \cite{schutzhub} that 
can still measure the parameters describing the Hubble flow 
over this interesting 
distance range. The method is interesting not 
only because it can determine parameters, but also because it is an example of 
a nonlinear statistical method whose errors improve much more rapidly 
with the number $N$ of samples than 
the usual $N^{1/2}$ associated with linear averaging, at least at first
when $N$ is small.

The idea is best illustrated for low-redshift measurements, where the 
goal is simply to determine one number, the Hubble constant. After 
understanding how this is done, we will see how it could be generalised
to larger redshifts. For each event, 
the detectors will produce an error box on the sky with 
a number of candidate clusters of galaxies in which the event may 
have occurred. The angular position of each candidate leads to a 
corresponding luminosity distance; measuring the mean redshift of 
each cluster then leads to a ``candidate'' value of the Hubble 
constant for that cluster. Each candidate cluster produces a 
candidate value. Most are wrong, but one of them should be 
the correct one. Now, if one has, say, ten such events, 
then one value of the Hubble parameter should appear in each 
set of candidates. As long as the number of candidate values is not
so large that the observational errors create a lot of overlap 
between false and true values, it should be possible to zero in 
on the correct value of the Hubble parameter and retrospectively 
identify the clusters in which the events occurred. With dozens of 
events, this method should be very efficient. 

This method is actually a one-dimensional version of the Hough 
transform method that was devised to analyse bubble-chamber photographs
in high-energy physics (see~\cite{leavers}). The tracks expected of particles were 
parametrised, and the number of bubbles on each possible track 
was counted. Real tracks would have a much larger number of bubbles
than random ones. The Hough transform is now being developed to search 
for unexpected pulsar signals in gravitational wave data~\cite{ht}.

For high-redshift objects, one could use the Hough transform to 
search for the best set of parameters for a cosmology. Appropriate
parameters might be the present Hubble constant, the value of 
$\Omega_{\text{matter}}$, and the value of $\Lambda$. With many 
tens of events, there should be enough statistics to find a set of
values that all the observations are consistent with. 
Of course, if by then the Hubble constant is well-enough known to 
determine the correct candidate from among the candidate clusters of galaxies, 
then this statistical method is not needed. 

Once the clusters 
containing the events are established, then the redshift and the 
luminosity distance can be used to calibrate the expansion of the 
universe. At this point the statistical errors of the measurements of
the large number of events will indeed reduce as $1/\sqrt{N}$. 
With data out to $z=1$, the parametrisation of the Hubble 
flow will be sensitive to the 
turnover, where the universe changed from deceleration to 
acceleration. We shall see that LISA can extend this method to 
redshifts of 4 or more.

\section{Cosmology with LISA}
LISA could measure a few mergers of massive black holes in 
the centres of galaxies each year. It will be sensitive to 
masses in the range $10^3$--$10^7\,M_\odot$. These observations
will give important insight into the processes that formed the 
black holes, into their population statistics, and into the 
role they played in galaxy formation. But here I wish to 
focus on the use of LISA to measure the Hubble flow itself.

For a typical 
system of two $10^6\,M_\odot$ black holes at $z=1$, 
LISA on its own will be able to determine the position to an accuracy
of about half a degree. The physical size of the error box
will be of the order of 40\,Mpc in all three dimensions, because
LISA's distance determination is also limited by the angular 
errors, as explained earlier. Thus, the error box is about 
1\% of the distance to the source.

This error box may contain a number of rich clusters, each 
with one or more candidate galaxies that show evidence of 
a past galaxy merger that could have led to the black hole merger.
We would like to identify the cluster in which the merger took place. 
There are at least three ways to do this.
\begin{enumerate}
\item If by the time LISA flies, the Hubble flow is known to an accuracy of 
better than 1\% out to redshifts of 4 or so, then this may 
assist identifying the cluster. One measures the redshifts of each 
of the candidates, and uses the angular positions of the candidates to 
determine from the LISA chirp signal what the luminosity distance is 
to that candidate. If the Hubble flow is known accurately as a 
function of luminosity distance, then the expected redshift can be compared
with the measured one, and if these do not coincide then the candidate 
can be rejected. If the expansion is known well enough, then the 
candidates may be narrowed down to just 1.
\item If the expansion is not known to this accuracy by the time LISA flies, 
then the statistical method described in the previous subsection 
could be brought to bear. This could work if there are of order ten 
events at high redshift over the mission 
lifetime of 10 years. The goal at this stage
would be to determine the Hubble flow accurately enough to identify the 
galaxies in which the events have taken place.
\item Failing both of these circumstances, it will be a challenge to
observers and astrophysicists to determine the galaxy in which the 
merger occurred by other means. Perhaps the morphology of the galaxy
is special in some way. The gradual in-spiral of the two massive
black holes transfers considerable energy to a number of stars in 
the core of the galaxy, and so it may be that the central bulge of 
this galaxy has a larger number of stars on nearly radial orbits 
than is normal. Or perhaps the two black holes maintained 
accretion disks, or even jets, until they came close enough to 
one another for tidal forces to disrupt them. The fossil jets 
may still be visible in the outer regions of the galaxy, and 
the gas of the accretion disks may have been expelled or shocked in 
a way that is observable for some time after the disruption.
\end{enumerate}

The identification of the merger galaxies has a large potential payoff.
The accurate angular 
position for each galaxy will provide a very accurate 
value of the luminosity distance, 
perhaps with errors smaller than 0.1\%, depending only on the SNR of
the detection. With the measured redshifts, then 
the measurements of the Hubble flow are limited
only by the proper velocities of the galaxies (inducing 
single-measurement uncertainties in the Hubble flow of 0.1\%). 
With a handful of 
merger events spread over redshifts out to, say, 4 or more, it should 
be possible to go well beyond $\Lambda$-cosmology models and test
quintessence and other models in which the pressure is not strictly 
equal to the negative of the energy density, and in which the 
density of dark energy/negative pressure is variable in time.

\section{Stochastic gravitational waves from the Big Bang}
Probably the most fundamental cosmological observation that gravitational wave 
detectors can make is of gravitational waves coming from the 
Big Bang. This is the gravitational analogue of the cosmic microwave 
background radiation, but with a key difference. Because 
gravitational waves couple so weakly to matter, they never 
thermalised. The non-thermal spectrum comes to us unchanged 
from whatever event(s) produced it. Using gravitational waves
we can see directly to the first fraction of a second after 
the Big Bang.

Unfortunately, what firm predictions exist for the amount of 
radiation that is produced by the Big Bang are discouraging. 
Gravitational waves should be created at some level by 
inflation in the same processes that produced the fluctuations
in energy density that led to galaxy formation, but the 
present energy density must be less than $10^{-13}$ of 
the closure density. The only observational limits are 
from the millisecond pulsar at frequencies of 1 cycle per
10 years, and from the requirement that the radiation not 
disturb nucleosynthesis. In both cases the limits require
$\Omega_{\text{gw}}$ to be smaller than about $10^{-6}$. 

Between the prediction of inflation and the observational limits
there is lots of room for other creative mechanisms, and 
many exist. Toy models of superstring cosmology can produce
tailor-made spectra with large amounts of radiation. 
Cosmic defects, phase transitions, and other unknown but 
not implausible physics can lead to radiation confined to 
certain wave-bands. Even brane-world cosmologies have 
the potential to produce radiation up to the nucleosynthesis 
limit at any frequency~\cite{hogan}.

Ground-based detectors can see this radiation best by 
cross-correlating the outputs of two nearby detectors. The best
suited are the two LIGO instruments. In the second generation they
may reach as low as , perhaps a little 
lower. LISA cannot cross-correlate its two independent interferometers
because they share a common arm and hence common noise. It can, 
however, internally calibrate its instrumental noise and thereby
identify any stochastic gravitational wave signal whose power is 
comparable to or larger than the instrumental noise. This is unlikely
to take it lower than $\Omega_{\text{gw}}\sim 10^{-10}$. I have 
proposed a variant of the LISA mission that could go down as 
low as $10^{-13}$, but this still does not reach the inflation 
prediction. A future LISA follow-on mission would be required to 
reach that level.

One of the problems with detecting a background from the Big Bang is
that there are astrophysical backgrounds of a more recent origin. This 
includes radiation from white-dwarf binary systems, ordinary binaries, 
close neutron-star binaries, and even small objects falling into 
massive black holes. There could be a window around 1\,Hz where 
the astrophysical backgrounds are weak enough to allow the cosmological 
background to dominate, but there are believed to be few other 
accessible windows~\cite{vecchioung}.

Only observations will tell us what is out there. LISA will certainly 
measure the compact white-dwarf binary background, which is expected to 
stand out well above the noise below 1\,mHz. LISA might also measure
backgrounds at higher frequencies. Whether LISA or the ground-based
detectors manages to see a cosmological background from fundamental 
physics near the Big Bang is one of the most unpredictable outcomes
of gravitational wave astronomy.

\section{Conclusions}
The astronomical community has waited a considerable time for 
gravitational wave detectors to realise their promise. The progress
has been steady but largely invisible until now. From next year, 
detections of some systems will be possible. But cosmological 
returns are likely to require another decade of development. 

The second-generation ground-based detectors should make the first 
impact on cosmology, providing values for the Hubble constant and
the acceleration of the universe that with an accuracy competitive 
with that of other methods. This will be a useful check on all methods.
The opening up of the low-frequency window by LISA after 2011 will 
bring much larger potential payoffs for cosmology. With some luck (or 
cleverness!), LISA could measure the deceleration/acceleration history 
of the universe with outstanding accuracy out to redshifts of 4 or earlier. 
To realise this promise, coordinated observations with telescopes in 
the optical/IR, X-ray, radio, and other bands will be essential.

%


\begin{thebibliography}{18.}
\addcontentsline{toc}{section}{References}

\bibitem{saulson} P. R. Saulson: {\em Fundamentals of Interferometric Gravitational
Wave Detectors} (World Scientific, Singapore, 1994)

\bibitem{hough} J Hough, S Rowan: Living Rev. Relativity {\bf 3}, 3 (2000). [Online article]: http://www.livingreviews.org/Articles/Volume3/2000-3hough/  (cited on 20 November 2001) 

\bibitem{MTW} C. W. Misner, K. S.  Thorne, J. L.  Wheeler: {\em Gravitation}
(Freeman \& Co., San Francisco, 1973)

\bibitem{firstcourse} B. F. Schutz: {\em A First Course in General Relativity}
(Cambridge University Press, Cambridge, 1985)

\bibitem{schutzwill} B. F. Schutz,  C.M. Will: `Gravitation and General Relativity'. In: {\em Encyclopedia of Applied Physics} {\bf 7}, 303-340 (1993)

\bibitem{schutzappphys} B F Schutz: `Gravitational Radiation'. In: {\em Encyclopedia of Astronomy and Astrophysics}  (Institute of Physics Publishing, Bristol, and Macmillan Publishers Ltd., London, 2000). Electronic version: gr-qc/0003069.

\bibitem{como} B. F. Schutz, F. Ricci:  `Gravitational Waves, Sources and Detectors', in: {\em Gravitational Waves}, ed. by I. Ciufolini, V. Gorini, U. Moschella, P. Fré (Institute of Physics 
Publishing, Londo, 2001).

\bibitem{lisa3} B. F. Schutz (ed): {\em Classical and Quantum Gravity}, {\bf 18}, Number 19 (7 October 2001).
  
\bibitem{kip300} K. S. Thorne: ``Gravitational Radiation''. In: {\em 300 Years of Gravitation},
 ed. by S. W. Hawking, W. Israel (Cambridge University Press, Cambridge, 1987), 
pp. 330--458

\bibitem{cutlerflanagan} C. Cutler, E. E. Flanagan: {\em Phys. Rev.} {\bf D49}, 2658 (1994)

\bibitem{cutler} A. Vecchio, C. Cutler, In: {\em Recent developments in theoretical and experimental general relativity, gravitation, and relativistic field theories, pt. B}, ed. by T. Piran (World Scientific, Singapore, 1999), pp. 1121--1123 

\bibitem{schutzhub} B.F. Schutz: {\em Nature} {\bf 323}, 310 (1986)

\bibitem{chernofffinn} L. S. Finn, D. F. Chernoff: {\em Phys. Rev.} {\bf D 47}, 2198 (1993)

\bibitem{heger} I. Baraffe, A. Heger, S. E. Woosley: {\em Astrophys. J.} {\bf 550}, 890 (2001)

\bibitem{markovic} D. Markovic: {\em Phys. Rev.} {\bf D 48}, 4738 (1993) 

\bibitem{chfinn} D F Chernoff, L. S. Finn: {\em Astrophys. J.} {\bf 411}, L5 (1993) 


\bibitem{finn} L. S. Finn: {\em Phys. Rev.} {\bf D 53}, 2878 (1996)

\bibitem{turner} Y. Wang, E. L. Turner: {\em Phys.Rev.} {\bf D 56}, 724 (1997) 

\bibitem{lorimer} D. R. Lorimer, E. P. J. van den Heuvel: {\em Mon. Not. Roy. astr. Soc.} {\bf 283}, L37 (1996)

\bibitem{port} S. F. Portegies Zwart, S. L. W. McMillan: {\em Astrophys. J.}
{\bf 528}, L17 (2000)

\bibitem{leavers} V. E. Leavers: {\em CVGIP: Image Understanding}, {\bf 58}, 2, 250 (1993)

\bibitem{ht} B. F. Schutz and M.-A. Papa In:  {\em Proceedings of Jan 1999 Moriond meeting "Gravitational Waves and Experimental Gravity"}, ed. by J. Trân Thanh Vân, J. Dumarchez, S. Reynaud, C.Salomon, S. Thorsett, J.Y. Vinet (Hanoi, 2000). Electronic version: gr-qc/9905018.

\bibitem{hogan} C. J. Hogan: {\em Class. Quant. Grav.} {\bf 18}, 4039 (2001).

\bibitem{vecchioung}  C. Ungarelli, A. Vecchio:
{\em Phys. Rev.} {\bf D 6306} 4030 (2001),  art. no. 064030




\end{thebibliography}
\end{document}